\documentclass[12pt]{elsarticle}
\usepackage[utf8]{inputenc}
\usepackage{graphicx}
\usepackage{subfigure}
\usepackage{array}
\usepackage{verbatim}

\usepackage{amssymb}
\usepackage{amsthm}
\usepackage{url}
\usepackage{amsmath}
\usepackage{subfigure}
\usepackage{textcomp}
\usepackage{xcolor}

\graphicspath{{figs/}}

\journal{XXXX}

\begin{document}

\begin{frontmatter}

\title{Logical circuits in colloids}
\author[a,b]{Nic Roberts}
\author[a]{Noushin Raeisi Kheirabadi}
\author[a]{Michail-Antisthenis Tsompanas}
\author[c,a]{Alessandro Chiolerio} 
\author[d]{Marco Crepaldi} 
\author[a]{Andrew Adamatzky}

\address[a]{Unconventional Computing Laboratory, UWE, Bristol, UK}
\address[b]{Department of Engineering and Technology, University of Huddersfield, UK}
\address[c]{Center for Bioinspired Soft Robotics, Istituto Italiano di Tecnologia,  Genova, Italy}
\address[d]{Electronic Design Laboratory, Istituto Italiano di Tecnologia,  Genova, Italy}

\begin{abstract}
Colloid-based computing devices offer remarkable fault tolerance and adapta\-bility to varying environmental conditions due to their amorphous structure. An intriguing observation is that a colloidal suspension of ZnO nanoparticles in DMSO exhibits reconfiguration when exposed to electrical stimulation and produces spikes of electrical potential in response. This study presents a novel laboratory prototype of a ZnO colloidal computer, showcasing its capability to implement various Boolean functions featuring two, four, and eight inputs.

During our experiments, we input binary strings into the colloid mixture, where a logical ``True" state is represented by an impulse of an electrical potential. In contrast, the absence of the electrical impulse denotes a logical ``False" state. The electrical responses of the colloid mixture are recorded, allowing us to extract truth tables from the recordings. Through this methodo\-logical approach, we demonstrate the successful implementation of a wide range of logical functions using colloidal mixtures.

We provide detailed distributions of the logical functions discovered and offer speculation on the potential impacts of our findings on future and emerging unconventional computing technologies. This research highlights the exciting possibilities of colloid-based computing and paves the way for further advancements.
\end{abstract}

\begin{keyword}
Unconventional computing, Colloids, Liquid computers, Liquid electronics, Liquid robotics
\end{keyword}

\end{frontmatter}

\section{Introduction}

A substance that cannot sustain shear stress when at rest is classified as a fluid: such stress necessarily produces a change in shape and the most remarkable dynamic phenomenon of flow. Specifically, a liquid is categorised as an incompressible fluid. Using liquids as computing devices can be traced back to documented evidence in papers discussing hydraulic algebraic machines \cite{emch1901two,gibb1914,frame1945machines}. In our recent comprehensive overview \cite{adamatzky2019brief}, we thoroughly examine various families of liquid computing devices. These include hydraulic  machine integrators, fluid mappers, fluid jets employed in fluidic logic devices to realise logical gates, liquid marble computers, and reaction-diffusion computers.

Several years ago, we developed the concept of the liquid cybernetic system ~\cite{chiolerio2017smart}, a colloidal autonomous system, which is a soft holonomic processor realising autolographic features~\cite{chiolerio2020liquid}. Further, these theoretical ideas of colloid computers were implemented into laboratory prototypes.
Our experiments conducted in controlled laboratory conditions also revealed the potential of ZnO colloid mixtures to function as electrical-analogue neurons, successfully implementing synaptic-like learning as described in \cite{kheirabadi2022learning}, as well as demonstrating the manifestation of Pavlovian reflexes \cite{kheirabadi2022pavlovian}. 
The experimental study presented in \cite{crepaldi2023experimental} showcases the classification capabilities of a Fe\textsubscript{3}O\textsubscript{4} water-based ferrofluid for digit recognition within an 8$\times$8 pixel dataset. Additionally, we demonstrated that this ferrofluid could be programmed using quasi-direct current signals and read in radio frequency mode. 

To thoroughly assess the computational capabilities of colloid computers more formally than previously explored, we embarked on a study to determine whether Boolean functions could be straightforwardly implemented in colloid mixtures. To achieve this, we adopted a theoretical approach outlined in \cite{adamatzky2017logical,adamatzky2020boolean}. This technique involves selecting a pair of input sites and systematically applying all possible combinations of inputs to these sites, where the electrical characteristics of the input signals represent logical values. The resulting outputs, represented by the electrical responses of the substrate, are recorded on a designated set of output sites.

This approach falls within the realm of reservoir computing \cite{verstraeten2007experimental,lukovsevivcius2009reservoir,dale2017reservoir,konkoli2018reservoir,dale2019substrate} and \textit{in materia} computing\cite{miller2002evolution,miller2014evolution,stepney2019co,miller2018materio,miller2019alchemy}, which are techniques used to analyse the computational properties of physical and biological substrates. By utilising these methodologies, we aim to gain deeper insights into the computational potential of colloid mixtures in a more rigorous and structured manner.

\section{Methods}
\label{methods}

\begin{figure}[!tbp]
    \centering
    \subfigure[]{\includegraphics[scale=0.65]{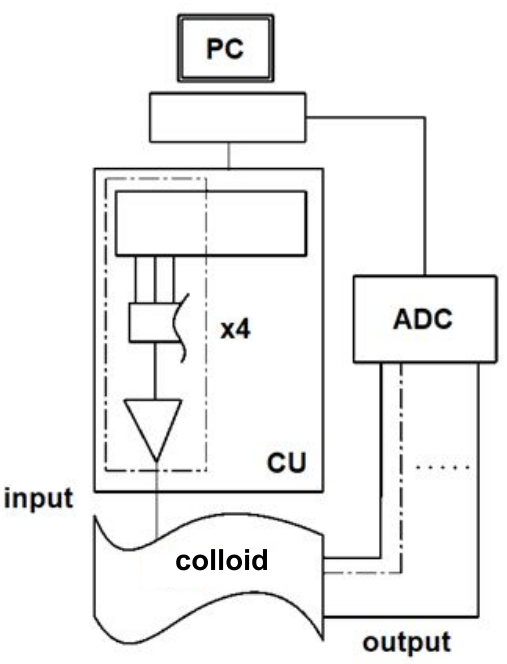}}
    \subfigure[]{\includegraphics[scale=0.12]{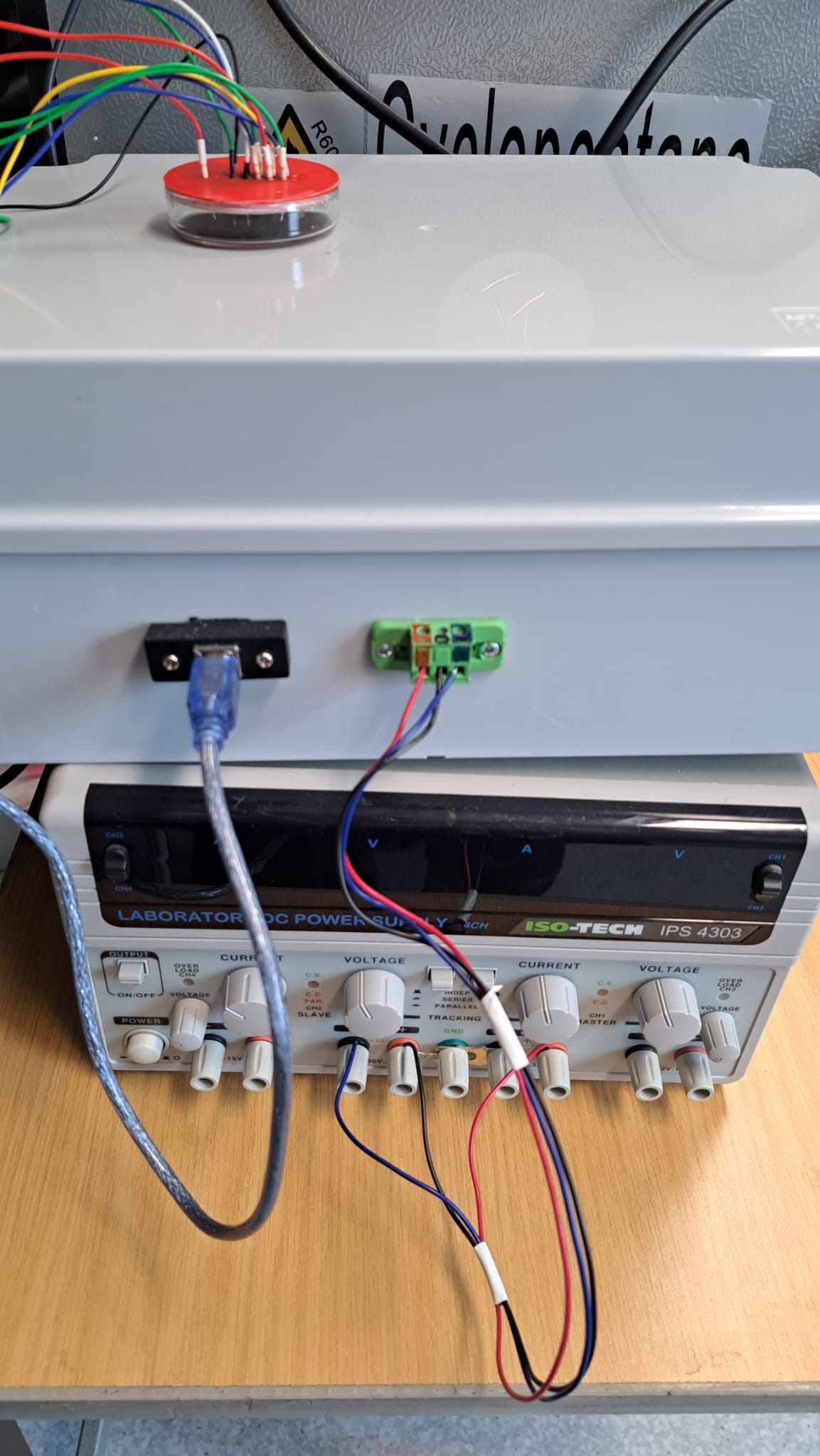}}
    \caption{Left: A scheme of the experiments. PC –- laptop for generating sequences; CU -- control unit, the dashed section is a breakdown of a single channel; ADC –- analog to digital converter. Right: experimental setup.}
    \label{fig:schematic}
\end{figure}

Zinc Oxide nanoparticles were purchased from US research nanomaterials. Sodium Dodecyl Sulphate (SDS) and Sodium Hydroxide (NaOH) were purchased from Merck. DMSO Pharmaceutical Grade 99.9\% were purchased from Fisher Scientific. A Millipore de-ionized water generating unit, model Essential, with a resistance of 15 Mohm cm, was used to create DIW in the lab. SDS was added to DIW and stirred to get a homogenous surfactant solution with a concentration of 0.22 wt\%. Under stirring, 2 ml of SDS solution and 1 ml of NaOH 10~M were added to the DMSO. The mixture was then treated with 1 mg ZnO nanoparticles while constantly stirring. The resulting dispersion concentration was kept constant at 0.11 mg/ml. For 30 minutes, the resultant suspension was placed in an ultrasonic bath. The stirring operation was then repeated for a few more hours to achieve a homogeneous dispersion of ZnO~\cite{anand2017role}.

The nanoparticle suspensions were characterized using field emission scanning electron microscopy (FEI Quanta 650 FESEM). In this study, the accelerating voltage was set to 10 kV, while the working distance was roughly 5 mm. The contrast and brightness of the photos were adjusted so that particles could be differentiated from the backdrop.
An Ultraviolet-visible spectrometer (Perkin Elmer Lambda XLS) was used to quantify sample absorbance at room temperature.
Dynamic Light Scattering (DLS) measurements were performed on a Zetasizer Nano ZS (1000 HS, Malvern Instrument Ltd., UK) to analyze the z-average hydrodynamic diameter.

The developed hardware could send sequences of 2, 4, and 8-bit strings to the colloid sample. The strings were encoded as step voltage inputs where -5~V denoted a logical '0' and 5~V a logical '1'. The hardware was based around an Arduino Mega 2560 (Elegoo, China) and a series of programmable signal generators, AD9833 (Analog, USA). 

To search for two-, four- and eight-input Boolean circuits, we used two, four, and eight-input electrodes, respectively. The input electrodes were 10~$\mu$m diameter platinum rods inserted into the colloid container with a separation of 5~mm between them. Data acquisition (DAQ) probes were placed in a parallel line, separated by 5~mm. There were 2 DAQ differential outputs from the sample container inputted to a Pico 24 (Pico Technology, UK) analogue-to-digital converter (ADC). The 3\textsuperscript{th} channel was used to pass a pulse to the ADC on every input state change. See Fig.~\ref{fig:schematic} for a schematic of the apparatus. There were a total of 138 repeats.

A sequence of two, four, and eight-bit strings counting up from binary
\textit{00} to \textit{11}, \textit{0000} to \textit{1111} and \textit{00000000} to \textit{11111111} with a state change every 15 seconds, were passed into the colloid. All 138 repeats of the experiment were done on the same colloid.
Samples from 2 channels were taken at 1~Hz over the whole duration of a given experimental run. Peaks for each channel were located for a set of 10 thresholds, from 100~mV to 600~mV with step 50~mV, for each input state, \textit{0000 to 1111}.

\section{Results}

\subsection{Colloid Structural Characteristics}

\begin{figure}[!tbp]
\centering
\includegraphics[width=\textwidth]{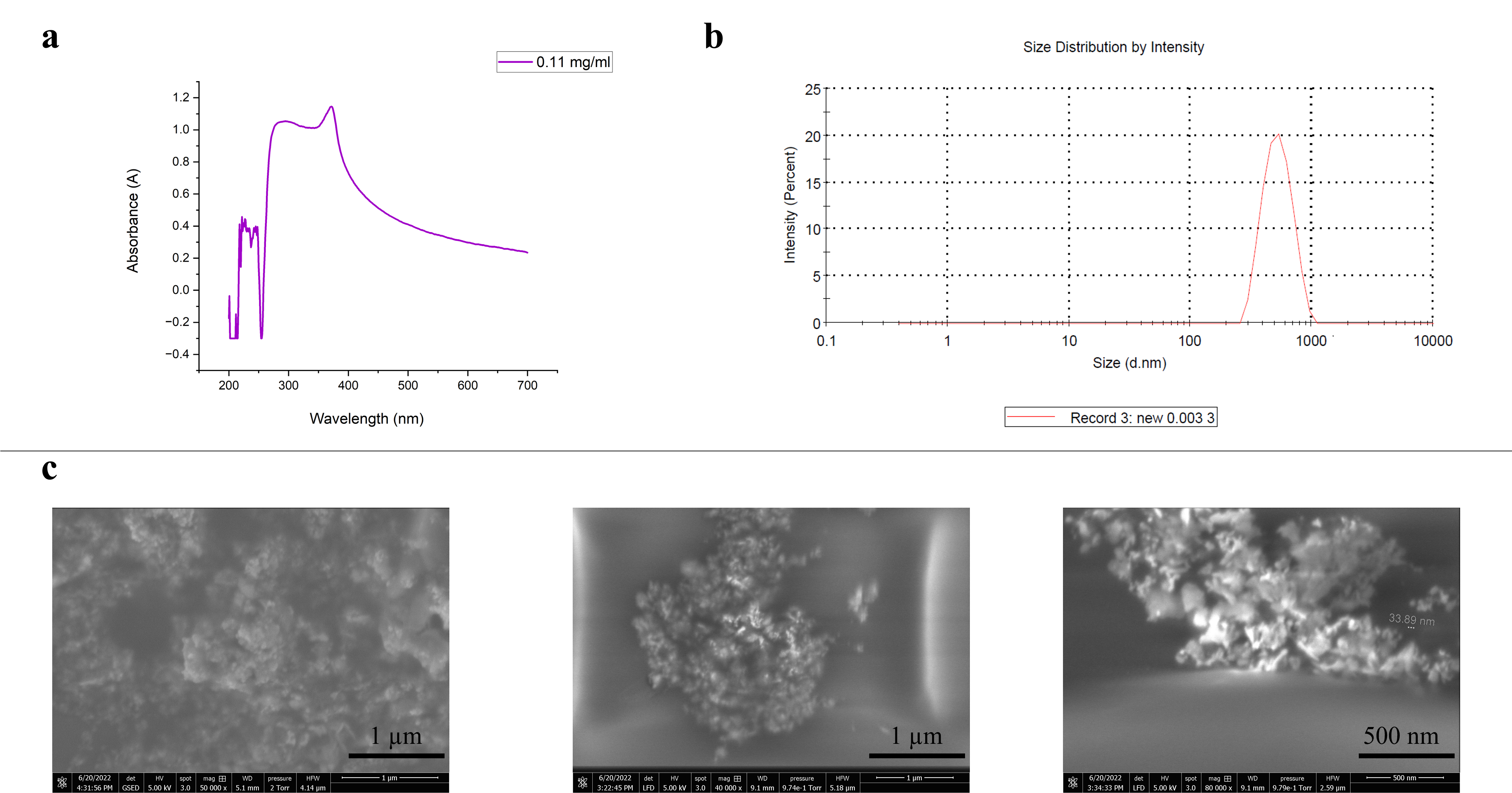}
\caption{Structural characterization of colloid used in experiments. (a)~UV-Visible spectra of the ZnO colloid. (b)~ Particle size distribution of ZnO nanoparticles in colloidal solution using dynamic light scattering. (c)~SEM images of drop-cast ZnO colloids on a Copper substrate in different magnifications.}
\label{fig: characterisation}
\end{figure}

The absorption spectrum of a ZnO colloid, with a concentration of 0.11 mg/ml, was measured at room temperature using UV-visible spectroscopy. The recorded spectrum covers a wavelength range of 200-700 nm. Figure~(\ref{fig: characterisation}, a) illustrates the UV-visible absorption spectrum plot. The spectrum shows a prominent peak at 372 nm, indicating hexagonal ZnO nanoparticles~\cite{pudukudy2015facile}. Comparing these findings with existing literature, there is a strong agreement with previous reports~\cite{reddy2011combustion,sun2011enhanced}.
The optical band gap was calculated using the following equation:
$$
E_g (eV) = hc/\lambda = 1240/\lambda
$$
In this equation, $E_g$ represents the optical band gap, h is Planck's constant, c is the speed of light, and $\lambda$ is the wavelength corresponding to the maximum absorption. The calculated value for the optical band gap is 3.35 eV, which aligns closely with previous findings from other sources~\cite{reddy2011combustion,baskoutas2010conventional,baskoutas2011transition}.

Dynamic Light Scattering (DLS) was utilized to characterize the ZnO nanoparticles in the colloid. Figure~(\ref{fig: characterisation}, b) displays the size distribution of these nanoparticles. A particle's average gyration (hydrodynamic) diameter is determined to be 496 nm, nearly 20 times the average diameter of an individual particle.
As an amphoteric oxide, ZnO undergoes hydrolysis when exposed to water, forming a hydroxide coating on its surface. This coating contributes to an increase in the hydrodynamic diameter of the particles~\cite{fatehah2014stability}.

A thin layer of ZnO colloid was prepared to analyse the particles' morphology and size using the FESEM (Field-Emission Scanning Electron Microscopy) technique. This was done by drop-casting a drop of ZnO particle suspension, with a concentration of 0.11 mg/ml, onto a Copper foil with a 100 $\mu$m thickness. The preparation was carried out at room temperature.

The FESEM results, as shown in Figure~(\ref{fig: characterisation}, c), reveal the occurrence of particle agglomeration during the sample preparation process. Due to the surface tension of the solvent as it evaporates, the FESEM observations rarely display individual, separated spheres. Instead, most of the ZnO spheres appear to be multilayered. This can be attributed to the increased liquid surface tension, which draws the nanoparticles closer and leads to their re-aggregation during the drying process~\cite{lu2018methodology}.

\subsection{Extracting Boolean Gates}

\begin{figure}[!tbp]
\centering
\includegraphics[width=\textwidth]{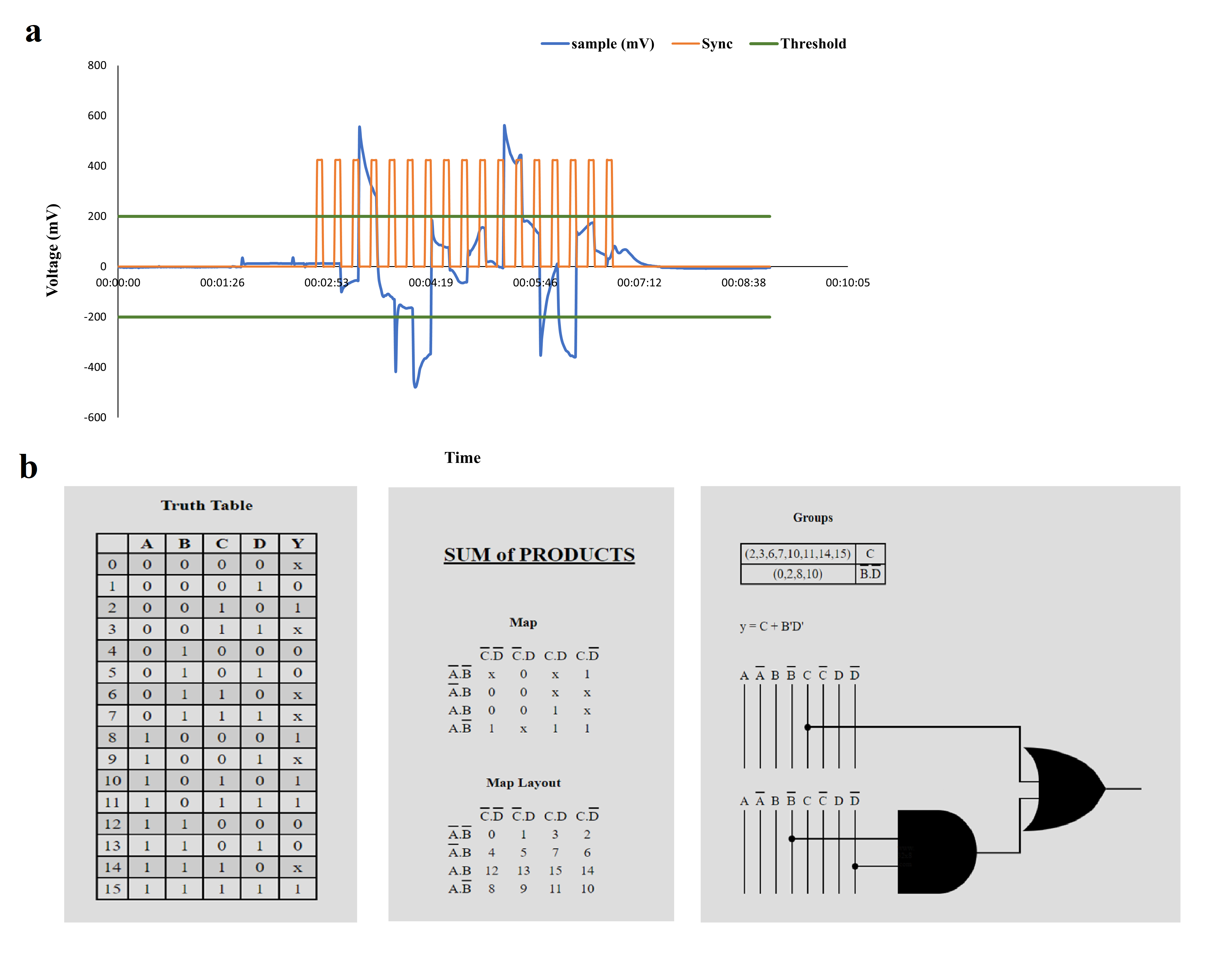}
\caption{Workflow example. a) The voltage measurements of the sample. The DAQ are in blue. The synchronisation signal is in orange, the threshold band is green.
 b) The resulting truth table and the extracted function.}
\label{fig:workflow}
\end{figure}

\begin{figure}[!tbp]
    \centering
\subfigure[]{\includegraphics[scale=0.41]{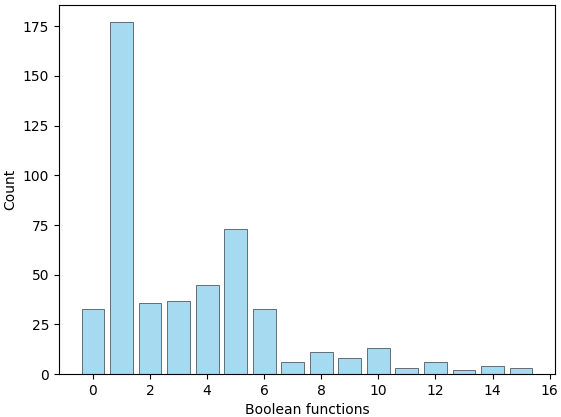}}
\subfigure[]{\includegraphics[scale=0.41]{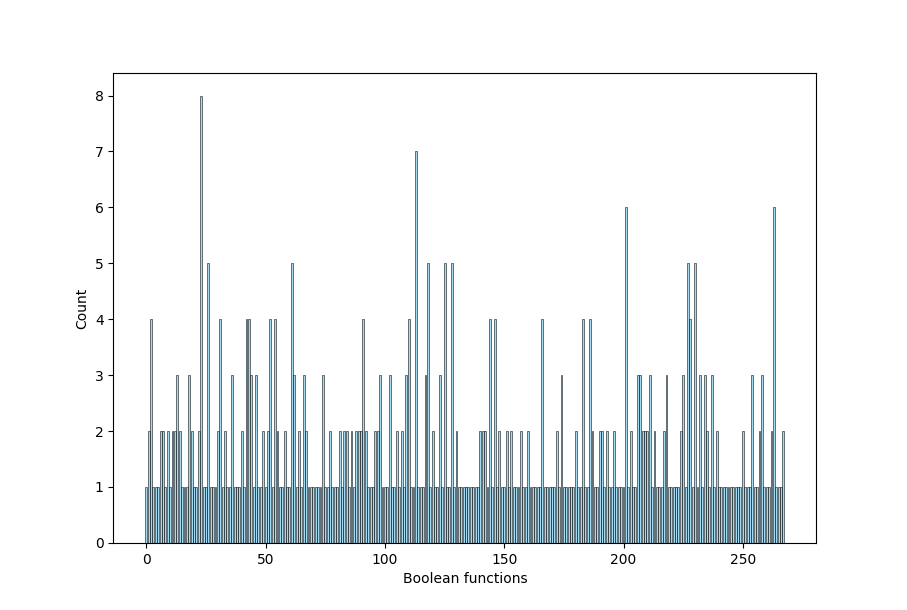}}
\subfigure[]{\includegraphics[scale=0.41]{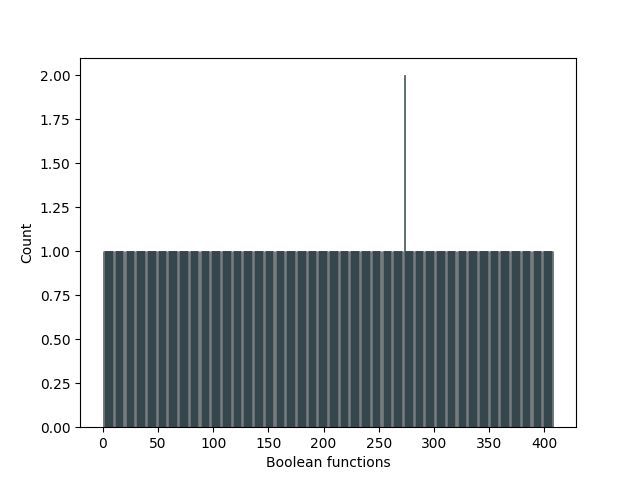}}
    \caption{Distribution of occurrences of two-input (a), four-input (b) and eight-input (c) Boolean gates discovered in laboratory experiments with colloids. The horizontal axis depicts Boolean functions in decimal codification. The vertical axis is the number of detected gates.}
    \label{fig:distributionOfgates}
\end{figure}

\begin{table}[!tbp]
    \centering
    \begin{tabular}{l|c}
Gate & $n$ \\ \hline
$\overline{A} + \overline{B}$ & 73 \\
$A + B$ & 45\\
$\overline{A} + B$ & 37\\
$A + \overline{B}$ & 33\\
$A\cdot B$ & 8\\
$B \cdot \overline{A}$ & 6\\
$(A \cdot \overline{B}) + (B \cdot \overline{A})$ & 4\\
$(A \cdot B) + (\overline{A} \cdot \overline{B})$ & 3\\
$A \cdot \overline{B}$ & 3\\
$\overline{A} \cdot \overline{B}$ & 2\\
    \end{tabular}
    \caption{Most commonly found two-input Boolean functions, $n$ is a frequency of the functions' discovery.}
    \label{tab:2inputgates}
\end{table}

\begin{table}[!tbp]
    \centering
    \begin{tabular}{l|c}
Gate & $n$ \\ \hline
$(A \cdot \overline{B}) + (B \cdot \overline{A} \cdot \overline{C}) + (B \cdot \overline{C} \cdot \overline{D})$ & 	 7 \\
$(C \cdot D \cdot \overline{B}) + (A \cdot \overline{B} \cdot \overline{D}) + (B \cdot \overline{A} \cdot \overline{D}) + (D \cdot \overline{A} \cdot \overline{C})$ & 	 6 \\
$(A \cdot \overline{B} \cdot \overline{D}) + (B \cdot \overline{A} \cdot \overline{C} \cdot \overline{D})$ & 	 6 \\
$(\overline{A} \cdot \overline{D}) + (A \cdot B \cdot C \cdot D) + (B \cdot \overline{A} \cdot \overline{C}) + (C \cdot \overline{A} \cdot \overline{B})$ & 	 5 \\
$(A \cdot \overline{B} \cdot \overline{D}) + (B \cdot \overline{A} \cdot \overline{C}) + (B \cdot \overline{C} \cdot \overline{D})$ & 	 5 \\
$A \cdot D \cdot \overline{B} \cdot \overline{C}$ & 	 5 \\
$A \cdot \overline{B} \cdot \overline{C} \cdot \overline{D}$ & 	 5 \\
$(B \cdot C \cdot D) + (B \cdot C \cdot \overline{A}) + (C \cdot D \cdot \overline{A}) + (A \cdot \overline{B} \cdot \overline{C} \cdot \overline{D})$ & 	 5 \\ 
$(A \cdot D \cdot \overline{B}) + (B \cdot D \cdot \overline{A}) + (A \cdot \overline{B} \cdot \overline{C}) + (B \cdot \overline{A} \cdot \overline{C}) + (D \cdot \overline{A} \cdot \overline{C})$ & 	 5 \\
$(D \cdot \overline{A}) + (D \cdot \overline{B}) + (B \cdot \overline{A} \cdot \overline{C})$ & 	 5 \\
    \end{tabular}
    \caption{Most commonly found four-input Boolean functions, $n$ is a frequency of the functions' discovery.}
    \label{tab:4inputgates}
\end{table}

\begin{figure}[!tbp]
    \centering
\subfigure[]{\includegraphics[scale=0.3]{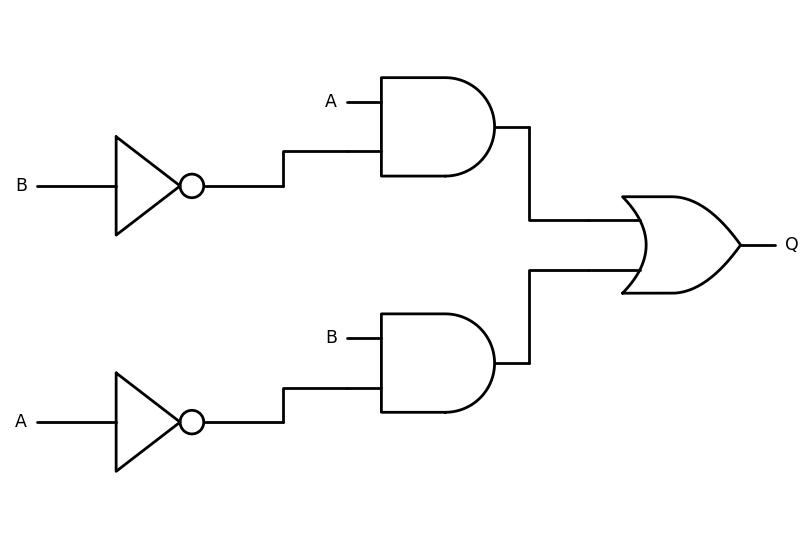}}
\subfigure[]{\includegraphics[scale=0.3]{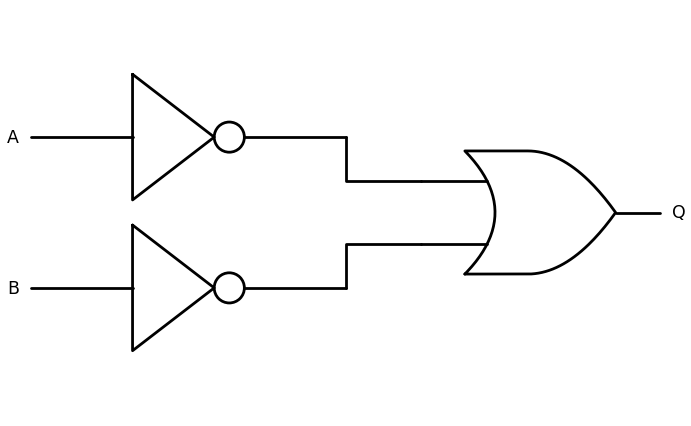}}
\subfigure[]{\includegraphics[scale=0.8]{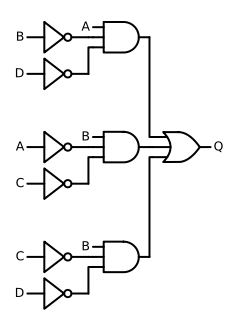}}
\subfigure[]{\includegraphics[scale=0.8]{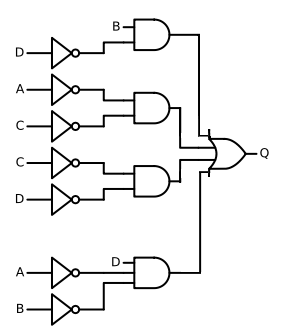}}
    \caption{Examples of logical circuits realised in (ab)~two-input and (cd)~four-input logical circuits. }
    \label{fig:exemplarcircuits}
\end{figure}

Boolean strings were extracted from the data, where a logic ‘1’ was noted for a channel if it had a peak outside the threshold band for a particular state. Otherwise, a value of ‘0’ was recorded, and the peak's polarity was not considered. 

The strings for each experimental repeat were stored in their respective Boolean table. To extract state graphs, a state/node was defined as the string of output values from each channel at each input state, and transitions/edges were defined as a change in the input state. This led to a total (500 + 470 + 410 = 1380) state graphs. The sum of products (SOP) Boolean functions were calculated for the output channel. For each repetition, we collected data and applied 10 thresholds, giving 1380 individual truth tables.

SOP extraction is depicted in Fig.~\ref{fig:workflow}. If a peak is discovered during an input state, it is considered a logical 1. The DAQ measurements are shown in blue. The synchronisation signal is in orange, indicating the state change. The threshold band is green, while peaks outside of it are marked with 'x'.
The resulting truth table is then reduced to the sum of products depicted in Fig.~\ref{fig:workflow}.

We have discovered a wide range of Boolean gates. Distributions of gates are shown in Fig.~\ref{fig:distributionOfgates}. Frequently found in experiments with two-input gates are shown in Tab~\ref{tab:2inputgates} and illustrated in terms of circuits in Fig.~\ref{fig:exemplarcircuits}ab. The most common gate is $\overline{A}+\overline{B}$, which is a {\sc NAND} gate, a logic gate producing an output that is false only if all its inputs are true; thus, its output is a complement to that of an {\sc AND} gate. The {\sc NAND} gate is followed by {\sc OR} gate and then by two {\sc NOT-AND} gates.

Most frequently four-input gates are shown in Tab.~\ref{tab:4inputgates} and illustrated with example circuits in Fig.~\ref{fig:exemplarcircuits}cd. 
The size of a Boolean circuit is the number of gates in the circuit. Amongst most frequent four-input circuits (Tab.~\ref{tab:4inputgates}) smallest circuits are 
$A \cdot overline{B} \cdot  \overline{C} \cdot D$ and 
$A \cdot \overline{B} \cdot \overline{C} \cdot \overline{D}$, and largest one is $(A \cdot D \cdot \overline{B}) + (B \cdot D \cdot \overline{A}) + (A \cdot \overline{B} \cdot \overline{C}) + (B \cdot \overline{A} \cdot \overline{C}) + (D \cdot \overline{A} \cdot \overline{C})$. 

The most frequent two-input and four-input gates are shown in Tab.~\ref{tab:2inputgates}. With regards to eight-input gates, all discovered gates are unique, i.e. have been measured just once, and the only following function has been found twice: 

$(A \cdot B \cdot F \cdot \overline{C} \cdot \overline{E})+(A \cdot D \cdot F \cdot \overline{C} \cdot \overline{E})+(A \cdot G \cdot H \cdot \overline{B} \cdot \overline{C})+(B \cdot D \cdot E \cdot \overline{A} \cdot \overline{F})+(B \cdot E \cdot H \cdot \overline{A} \cdot \overline{C})+(C \cdot D \cdot E \cdot \overline{B} \cdot \overline{F})+(D \cdot E \cdot H \cdot \overline{B} \cdot \overline{G})+(D \cdot F \cdot H \cdot \overline{A} \cdot \overline{B})+(B \cdot C \cdot D \cdot F \cdot G \cdot \overline{H})+(B \cdot C \cdot D \cdot G \cdot H \cdot \overline{F})+(B \cdot E \cdot \overline{A} \cdot \overline{G} \cdot \overline{H})+(C \cdot F \cdot \overline{B} \cdot \overline{G} \cdot \overline{H})+(E \cdot H \cdot \overline{A} \cdot \overline{C} \cdot \overline{F})+(F \cdot H \cdot \overline{B} \cdot \overline{D} \cdot \overline{E})+(A \cdot C \cdot E \cdot G \cdot \overline{B} \cdot \overline{D})+(A \cdot E \cdot F \cdot G \cdot \overline{C} \cdot \overline{D})+(B \cdot D \cdot E \cdot G \cdot \overline{C} \cdot \overline{F})+(B \cdot E \cdot F \cdot G \cdot \overline{A} \cdot \overline{D})+(C \cdot F \cdot G \cdot H \cdot \overline{D} \cdot \overline{E})+(A \cdot C \cdot D \cdot E \cdot F \cdot H \cdot \overline{G})+(B \cdot \overline{C} \cdot \overline{E} \cdot \overline{G} \cdot \overline{H})+(C \cdot \overline{B} \cdot \overline{D} \cdot \overline{F} \cdot \overline{G})+(D \cdot \overline{C} \cdot \overline{E} \cdot \overline{F} \cdot \overline{H})+(E \cdot \overline{A} \cdot \overline{B} \cdot \overline{D} \cdot \overline{H})+(E \cdot \overline{B} \cdot \overline{D} \cdot \overline{G} \cdot \overline{H})+(B \cdot C \cdot D \cdot \overline{A} \cdot \overline{E} \cdot \overline{G})+(B \cdot D \cdot F \cdot \overline{C} \cdot \overline{G} \cdot \overline{H})+(B \cdot D \cdot G \cdot \overline{A} \cdot \overline{C} \cdot \overline{E})+(B \cdot E \cdot H \cdot \overline{C} \cdot \overline{F} \cdot \overline{G})+(C \cdot D \cdot G \cdot \overline{B} \cdot \overline{E} \cdot \overline{H})+(C \cdot E \cdot F \cdot \overline{D} \cdot \overline{G} \cdot \overline{H})+(C \cdot G \cdot H \cdot \overline{A} \cdot \overline{E} \cdot \overline{F})+(D \cdot E \cdot F \cdot \overline{A} \cdot \overline{B} \cdot \overline{C})+(B \cdot D \cdot \overline{E} \cdot \overline{F} \cdot \overline{G} \cdot \overline{H})+(B \cdot F \cdot \overline{A} \cdot \overline{D} \cdot \overline{E} \cdot \overline{G})+(C \cdot D \cdot \overline{A} \cdot \overline{B} \cdot \overline{E} \cdot \overline{H})+(C \cdot H \cdot \overline{D} \cdot \overline{E} \cdot \overline{F} \cdot \overline{G})+(A \cdot C \cdot E \cdot G \cdot \overline{D} \cdot \overline{F} \cdot \overline{H})+(A \cdot \overline{B} \cdot \overline{C} \cdot \overline{F} \cdot \overline{G} \cdot \overline{H})+(D \cdot \overline{A} \cdot \overline{B} \cdot \overline{C} \cdot \overline{E} \cdot \overline{G})+(G \cdot \overline{A} \cdot \overline{C} \cdot \overline{D} \cdot \overline{E} \cdot \overline{H})$

\section{Discussion}

Our laboratory experiments successfully demonstrated the feasibility of implementing a wide range of many-input logical gates within a colloid mixture comprising ZnO nanoparticles. The discovered two-input gates exhibit functional completeness, enabling the implementation of arbitrary Boolean functions. Notably, the four- and eight-input functions discovered showcase a remarkable level of non-linearity, suggesting that the dynamical behaviour of colloid-based logical devices could be characterized by multiple attractors and bifurcation points, more than features such as resistive switching which was already observed~\cite{RSCA2016}.

Looking ahead, our future research direction could focus on cascading the colloid droplets to construct multi-level logical circuits. Additionally, we aim to develop protocols for programming dynamical logical circuits within the colloid droplets. Leveraging on the observed multistability that is the presence of multiple attractors, a possible blue sky objective would regard the implementation of a sequential computing machine, which, similarly to solid-state computers, can be programmed and can execute instructions. Our laboratory experiments have already demonstrated the fundamental assumptions needed to reach this goal, showcasing the possibility of imple\-menting in-memory computing with ferrofluids, therefore showing the coexis\-tence of memorising and computing capabilities. For the particular case studied here, the presence of multiple Boolean transfer functions inherently suggests the existence of a pre-built program in the colloid, here, a function of threshold, but in general, a function of other parameters, including time. A possible solution to achieve functionalities similar to a microprogrammed solid-state computer could regard modulating input signals to convey an equivalent inline and just-in-time executed program. Further studies can then focus on the stimulations' meaning and timing and the feasible techniques for their modulation. This concept, \textit{inter alia}, further overlaps with the field of neuromorphic computing because inputs can degenerate into spikes for low-duty cycles. 
These advancements would enhance the capabilities and expand the potential applications of colloid-based computing systems.

\section{Acknowledgement}

This project has received funding from the European Innovation Council And SMEs Executive Agency (EISMEA) under grant agreement No. 964388.


\end{document}